# Systematic Quantum Mechanical Region Determination in QM/MM Simulation

Maria Karelina[1] and Heather J. Kulik[1],*

[1]*Department of Chemical Engineering, Massachusetts Institute of Technology, Cambridge, MA 02139*

ABSTRACT: Hybrid quantum mechanical-molecular mechanical (QM/MM) simulations are widely used in enzyme simulation. Over ten convergence studies of QM/MM methods have revealed over the past several years that key energetic and structural properties approach asymptotic limits with only very large (ca. 500-1000 atom) QM regions. This slow convergence has been observed to be due in part to significant charge transfer between the core active site and surrounding protein environment, which cannot be addressed by improvement of MM force fields or the embedding method employed within QM/MM. Given this slow convergence, it becomes essential to identify strategies for the most atom-economical determination of optimal QM regions and to gain insight into the crucial interactions captured only in large QM regions. Here, we extend and develop two methods for quantitative determination of QM regions. First, in the charge shift analysis (CSA) method, we probe the reorganization of electron density when core active site residues are removed completely, as determined by large-QM region QM/MM calculations. Second, we introduce the highly-parallelizable Fukui shift analysis (FSA), which identifies how core/substrate frontier states are altered by the presence of an additional QM residue on smaller initial QM regions. We demonstrate that the FSA and CSA approaches are complementary and consistent on three test case enzymes: catechol O-methyltransferase, cytochrome P450cam, and hen eggwhite lysozyme. We also introduce validation strategies and test sensitivities of the two methods to geometric structure, basis set size, and electronic structure methodology. Both methods represent promising approaches for the systematic, unbiased determination of quantum mechanical effects in enzymes and large systems that necessitate multi-scale modeling.

## 1. Introduction

A firm understanding of how enzymes facilitate chemical reactions is crucial, and atomistic simulations of enzymes[1] can provide valuable insight into the source of enzymatic rate enhancements. In order to enable a sufficient balance in accuracy to describe chemical rearrangements and catalytic enhancement with the low computational cost needed to enable sampling, a multilevel approach is employed,[2-10] wherein the region of primary interest is treated quantum mechanically (QM), while the surrounding portion of the enzyme is described with an empirical molecular mechanics (MM) model. Quantum mechanical descriptions are essential in modeling enzyme catalysis for their ability to describe charge transfer, polarization, and bond rearrangement. Due to computational limitations, typical QM region sizes have been on the order of tens of atoms (i.e. ligands and a few direct residues).[10-12] These small QM regions have motivated method development[8, 13-22] to minimize QM/MM boundary effects and to evaluate[23] how advanced, polarizable,[24-25] force field treatments may improve QM/MM descriptions. However, the requirement to treat crucial[26-27] charge transfer across the QM/MM boundary suggests that boundary-effect minimization and force field adjustment may be insufficient to address the shortcomings of small QM/MM calculations.

Recent advances[26, 28-35] in computational efficiency now enable fully *ab initio*, quantum chemical simulation of polypeptides[28] as well as QM/MM treatments of enzymes with large (> 100 atoms) QM regions. Harnessing these computational advances, numerous researchers[36-45] have carried out studies to identify how quickly properties reach asymptotic limits as QM regions are enlarged in QM/MM calculations. The resulting studies have revealed an exceptionally slow approach to asymptotic limits for: NMR shieldings,[36-37] solvation effects,[38] barrier heights,[39-41] forces[42], excitation energies,[43] partial charges,[44] and redox potentials[45] with most properties predicted to converge at no fewer than 500 atoms (e.g., barrier heights[39-41] and forces[42]) but more typically on the order of 1000 atoms (e.g., NMR shieldings[36-37] and partial charges[41, 44]). Some of us recently studied the convergence of properties[41], i.e., structures, barrier heights, reaction enthalpies, and partial charges, on the enzyme catechol O-methyltransferase[46] (COMT) with increasing QM region size in QM/MM calculations. We again observed slow convergence to asymptotic limits on the order of 500 or more atoms, and we also identified large-scale quantum-mechanical effects to be essential[41, 47] to reproducing experimental structural properties and identifying sources of enzymatic rate enhancements.

Although it has been possible to carry out single point energies of very large biomolecules for some time,[48-54] quantitative enzyme mechanism study requires thousands of such single point energies. Thus, equally important to determining the approach to asymptotic limit in QM/MM studies as larger QM regions are used is the development of atom-economical and systematic strategies to achieve accurate descriptions of the electronic environment around a central region of interest (e.g., reacting active site residues). Systematic determination strategies would also eliminate any error due to limitations in chemical intuition or prior knowledge about the reaction mechanism from experiments. Some schemes have been suggested for constructing large QM regions including free energy perturbation analysis[55] or charge deletion analysis[39], both of which rely on determining the effect on a small QM region of changing an aspect of the MM environment. Some of us[41] recently suggested the development of an alternative methodology, charge shift analysis (CSA), to enable systematic QM region determination. Namely, we observed that residues that exhibited a redistribution of charge density when the core active site substrates were removed could be used to construct an electronically complete QM region consistent with asymptotically converged QM region predictions on COMT.

The potential limitations of the CSA method as it was first introduced were that i) it was carried out based on roughly 1000-atom QM region QM/MM calculations, which may remain prohibitive for some combinations of quantum chemistry software and hardware, and ii) only complete removal of the substrates was demonstrated, which is impractical for many enzymes where the core active site is defined by covalently bound protein residues or substrates. In this work, we thus develop and test the CSA method on covalently-bound substrates and validate our earlier observations[41]. We also introduce a new method that does not require large-scale electronic structure calculations for QM region determination and is insensitive to how the core active site is defined (i.e., if covalently bound residues are included). Although some efforts have been made to suggest general[39, 56-59] or system-specific[45] QM region determination, QM regions in QM/MM methodology are still largely determined by trial and error. To emphasize the systematic nature of the two approaches introduced here, we also introduce validation methodology for assessing the results of these two new methods. The outline of the rest of this work is as follows. In Sec. 2, we introduce two strategies for systematic QM/MM region determination. In Sec. 3, we provide the computational details of our study. In Sec. 4, we present Results and Discussion of applying and validating our new methodology on three diverse enzymes. Finally, in Sec. 5, we provide our conclusions.

## 2. Approach

## 2a. Charge Shift Analysis

Some of us first introduced[41] the idea that the electronic environment around a series of core active site residues may be probed by identifying how the density is redistributed when these core residues are rigidly removed. This approach has parallels with the rigid binding interaction energy approach of MMPBSA/MMGBSA[60] and energy decomposition analysis[61] in the force field and electronic structure communities, respectively. In this strategy, we start from a very large QM region selected by radial distance cutoffs from a central region of interest in a QM/MM calculation. This QM region size should meet or exceed asymptotically converged QM region sizes, which are typically up to 1000 atoms in size[23, 36, 39, 41-42, 44-45, 55, 58-59, 62-64]. Here, we choose 900 to 1000 atoms in the holoenzyme QM region in these QM/MM calculations, due to the increasingly routine application of electronic structure methods to these larger system sizes. Next, we remove the core substrate residues and recalculate all electronic structure properties of what we refer to as the apoenzyme. In our previous work[41], the core active site consisted of substrate residues that were not covalently bound to the active site and could be readily removed (Figure 1). For many enzymes, the core active site instead consists of both protein residues or other covalently linked cofactors and non-covalently bound species. In that case, we propose generation of the "apoenzyme" by alanine mutagenesis of all "core" protein residues or covalently bound substrates and elimination of non-covalently bound substrate molecules (e.g., natural products, DNA, surrounding lipids, water) (Figure 1). For non-protein simulations, the alanine mutagenesis step is instead removal and methyl-capping of the covalently linked active site region. Methyl-capping is chosen here to avoid introducing bond polarization effects due to low hydrogen atom electronegativity, e.g. with replacement of the sidechain with a hydrogen atom as in glycine mutagenesis or complete removal of the residue and capping of neighboring residues with link atoms. A comparison to alternative capping strategies is also considered in this work (see Results and Discussion).

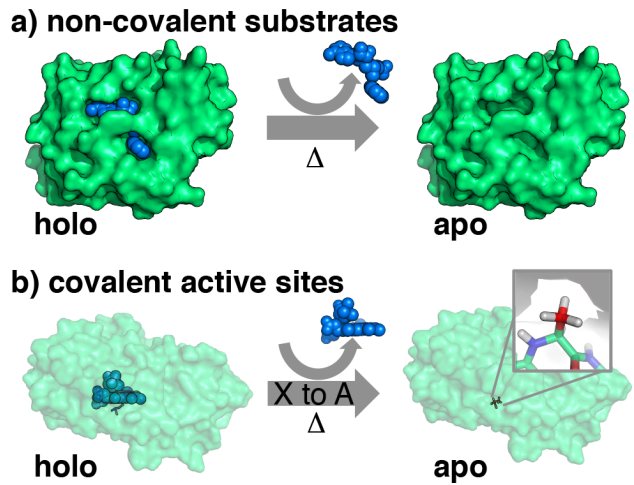

**Figure 1.** Schematic of the CSA approach for a) non-covalently bound substrates and b) covalently bound active site residues. The protein surface is shown in green and the active site is shown in blue spheres and sticks. In the case of (b), the covalently bound residues are also mutated to Ala (shown inset, in red).

In order to evaluate the results of the CSA method, we sum over the partial charges ($q$) of the $N_{at}$ atoms in each QM residue (RES):

$$q_{\text{RES}} = \sum_i^{N_{at} \in \text{RES}} q_i \qquad (1)$$

These sums can be carried out with or without the inclusion of link atoms within each residue definition, and we recommend exclusion of link atoms for numerical stability in comparisons. The charge shift may then be calculated as the difference of the summed-over-residue partial charge of the non-core-active-site residue between the complete enzyme (holo) and when the active site residues are removed or mutated (apo):

$$\Delta q_{\text{RES}} = q_{\text{RES}}^{\text{holo}} - q_{\text{RES}}^{\text{apo}} \ . \qquad (2)$$

If the partial charge difference on a residue, $\Delta q_{\text{RES}}$, exceeds a recommended cutoff of about |0.04-0.05 e|, based on uncertainty in partial charge estimates, then this residue's electronic properties

are deemed affected by the presence or absence of the core active site. We have selected this threshold based on observations that numerical noise in partial charge differences can produce small (ca. 0.01-0.02 e) differences in by-residue sums between simulations (see tabulated values in the Supporting Information). This threshold works well for the three proteins studied in this work, but one may confirm an appropriate threshold by sorting charge shifts by magnitude and identifying where the values fall off to a background level. Additionally, validation methodology introduced later in this work may be useful for testing the importance of residues that are near the cutoff value (see Results and Discussion).

Importantly, this approach incorporates many-body effects, detecting charge redistribution in a residue due to electronic structure changes on another residue in addition to direct changes due to the removal of substrates. Due to the nature by which the electronic structure calculation is carried out, the possible residues identified through this method must have already been included in the original large-QM simulation, motivating the 900-1000 atom QM regions selected in this work. Nevertheless, a smaller original QM calculation could be employed if one wished to select from a shorter list of residues, albeit with some potential disadvantages, as outlined in the Results and Discussion. Because partial charges are summed over residues, and their differences are calculated under perturbation of active site removal, the method may be expected to be relatively basis set insensitive (see also Results and Discussion). Nevertheless, the use of exchange-correlation functionals that can reliably treat charge transfer, i.e., range-separated hybrids becomes necessary. Additionally, in principle, this approach requires only two calculations – a single point each for the holoenzyme and the apoenzyme, unless structural averaging is motivated (see Results and Discussion).

We note that the CSA approach is related to but diverges from the charge deletion analysis (CDA) method[39, 57, 65], wherein residue effects on relative QM energies are probed by deletion of associated MM point charges:

$$\Delta\Delta E(\text{RES}) = \Delta E\big|_{\text{QM/MM,MM(RES)}} - \Delta E\big|_{\text{QM/MM,MM(RES)}\to 0} \quad . \tag{3}$$

In contrast, the CSA method assumes that properties may be affected by MM point charges without the need to incorporate them into a QM region unless they also impact electronic properties. Additionally, CDA shows marked size-dependence (i.e., CDA on different QM region sizes gives very different results[39, 56]). The CSA method would only depend on the QM region size in one of two ways: i) detected residues must have been included in the original QM calculation, and ii) very small CSA QM region calculations, as with standard small QM calculations, may introduce boundary effects. We provide further comparison of the two methods in the Results and Discussion.

**2b. Fukui Shift Analysis**

Thus far, we have introduced the CSA method, which necessitates the direct calculation of large-QM QM/MM electronic structure calculations with QM regions between 500 and 1000 atoms in size. In order to enable high-throughput, parallelizable QM/MM region determination, we have developed an alternative approach to probe the influence of protein residues on electronic properties at enzyme active sites. Here, we focus on extracting residue-specific influences on core active site properties. We scan through every residue in the protein and sequentially add that residue to a model that also includes core active site residues (Figure 2).

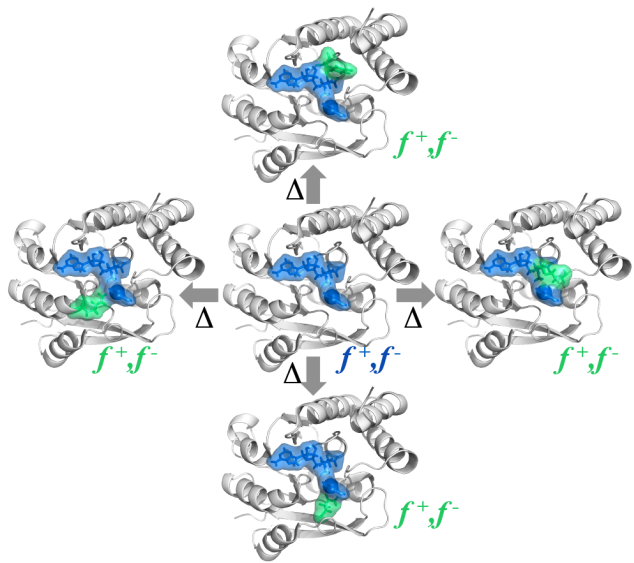

**Figure 2.** Schematic of the FSA approach showing core model (center) with core residues and other QM residues in sticks and represented by a surface. The Fukui shift is computed by calculating the condensed Fukui functions, adding a single residue from the protein shown in green sticks, as indicated schematically by the top, left, right, and bottom proteins and calculating the change in the condensed Fukui function.

We then evaluate how that additional residue alters the local properties of the core active site. Rather than focusing on charge transfer alone, which may be sensitive to our use of a single additional residue, we invoke models of reactivity widely employed in the context of conceptual density functional theory. Namely, the Fukui function[66] is a three-dimensional representation of the density gained or lost on electron addition or removal. Typically, the Fukui function is obtained at integer electrons, either from electron addition:

$$f_+(\mathbf{r}) = \rho_{N+1}(\mathbf{r}) - \rho_N(\mathbf{r}) \tag{4}$$

or removal:

$$f_-(\mathbf{r}) = \rho_N(\mathbf{r}) - \rho_{N-1}(\mathbf{r}) \ . \tag{5}$$

It is more convenient to work directly with the condensed Fukui functions[67], which are evaluated as the atomic contribution to the overall Fukui function with electron addition or removal, respectively, as partitioned through any calculated partial charges (i.e., $q_i$ on atom $i$):

$$f_+^i = q_i(N+1) - q_i(N) \tag{6}$$

and

$$f_-^i = q_i(N) - q_i(N-1) \; . \tag{7}$$

Condensed Fukui functions have been applied to biological systems previously[68], and the placement of frontier states has been used to identify enzyme active sites[69] in conjunction with semi-empirical methods. Numerical robustness of this approach is aided by the fact that we employ Voronoi deformation density (VDD) partial charges[70], which exhibit relatively low basis set sensitivity due to the use of real-space partitioning (see sec. 3). Furthermore, we focus on a fragment-based condensed Fukui function wherein we sum all components of the Fukui function over an entire substrate molecule or protein residue in the core active site, obtained as:

$$f_+^{\text{RES}} = \sum_i^{N_{at} \in \text{RES}} q_i(N+1) - q_i(N) \tag{8}$$

or

$$f_-^{\text{RES}} = \sum_i^{N_{at} \in \text{RES}} q_i(N) - q_i(N-1) \; . \tag{9}$$

We evaluate how condensed Fukui functions on essential active site residues (ASR) vary upon inclusion of a single additional protein residue. We choose as the reference values the median of all Fukui results on distant, neutral residues, which generally have a narrow distribution of Fukui functions that correspond to no influence of the residue over the active site. We then compute an effect of a residue to be the root sum squared (RSS) difference in by-residue Fukui functions (RSS($f$)) with respect to that median value:

$$\text{RSS}(f) = \sqrt{\sum_{j}^{\text{ASR}} \left( \left( f_{+}^{j} - \tilde{f}_{+}^{j} \right)^2 + \left( f_{-}^{j} - \tilde{f}_{-}^{j} \right)^2 \right)} \ . \tag{10}$$

Because each calculation consists of at most one residue in addition to the active site residues, it becomes straightforward to rapidly scan through all residues in the protein in a highly parallelized fashion. The condensed Fukui function sums used in FSA are expected to be strongly reliant upon the use of an exchange-correlation functional that preserves band gaps and models charge transfer accurately. Some active-site/residue pairings particularly challenge the electronic structure method with non-contiguous QM regions, and the importance of asymptotically correct range-separated hybrids to avoid false positives becomes apparent (see Results and Discussion). Potential challenges with this method include choice of spin for the added or removed electron reference, much as within other applications of Fukui functions. If more residues are incorporated into the model through which residues are scanned, some have suggested[68] adding or removing more than one electron when computing the Fukui reference. As a general approach, this method assumes the impact of each residue may be evaluated one at a time, and we will provide a comparison between the CSA and FSA methods in the Results and Discussion to identify any cases where this assumption fails.

We also note that for both CSA and FSA, beyond partial charge sums or Fukui functions, it may be desirable to evaluate electron density redistribution in terms of local dipole moments or atomic polarizabilities[71]. Nevertheless, polarizability and changes in dipole moments could potentially be captured by a suitably accurate polarizable force field, whereas charge transfer cannot be modeled across the QM/MM boundary.

**3. Computational Details**

*Protein structure preparation*. A general protocol for all proteins simulated in this work was as follows. The protein crystal structures were obtained from the protein databank for hen eggwhite lysozyme (HEWL, PDB ID: 2VB1[72]), cytochrome P450cam (P450cam, PDB ID: 1DZ9[73]), and catechol O-methyltransferase (COMT, PDB ID: 3BWM[74]). Structures of the active sites of these three prepared proteins are shown in Figure 3. The charge state of the holoenzymes, or apoenzymes where applicable (i.e., P450cam and COMT), was assigned using the H++ webserver[75-78] assuming a pH of 7.0 with all other defaults applied. As H++ removes all nonstandard residues, residues in the active site adjacent to cofactors or substrates were manually assigned protonation states based on previous literature results[23, 42] (see Supporting Information Tables S1-S3). The output of H++ was used as the starting point for subsequent topology and coordinate preparation using the AMBER[79] tleap utility prior to classical molecular dynamics (MD) and QM/MM simulation with AMBER[79].

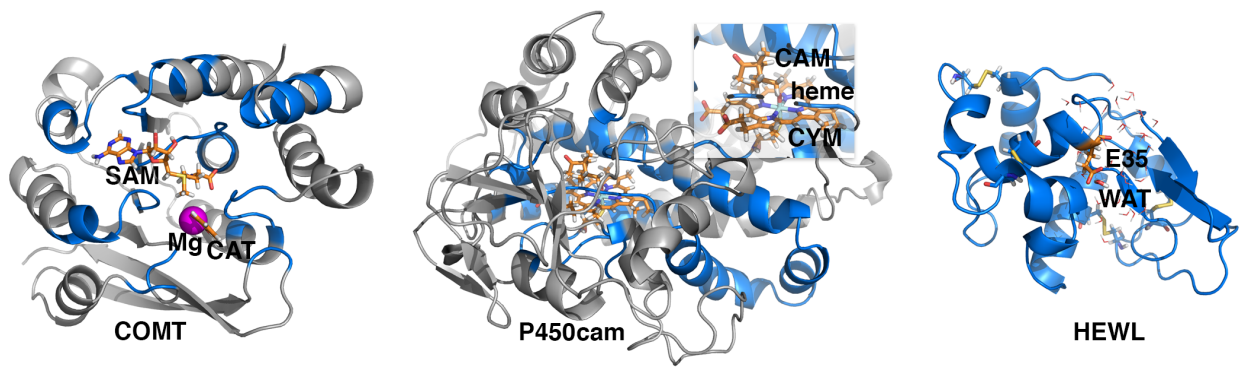

**Figure 3.** Active sites of proteins studied in this work: (left) COMT, (center) P450cam, and (right) HEWL. Core residues have been shown as sticks in orange and annotated. The QM residues in CSA calculations are shown in blue, and the rest of the protein, where relevant, is shown in gray.

For HEWL, four disulfide bridges are present, which were manually bonded when finalizing preparation in tleap. Protein residues were described by the AMBER ff14SB[80] force field, which is derived from the ff99SB[81] force field with updates to backbone torsional parameters. For nonstandard residues in COMT and P450cam, we employ the generalized AMBER force field (GAFF)[82] with partial charges assigned from restrained electrostatic potential (RESP) charges[83] obtained with GAMESS-US[84] at the Hartree-Fock level using a 6-31G*[85] basis set, as implemented by the R.E.D.S. web server[86-88]. Resolved water molecules in the crystal structures were preserved, whereas any crystallizing agents were removed. The $Mg^{2+}$ force field parameters used in COMT simulation were obtained from Ref. [89], as validated in some of our previous work[90]. Each protein was solvated in a periodic rectangular prism box with at least a 10 Å buffer of TIP3P[91] water and neutralized with either $Na^+$ or $Cl^-$ counterions. The full simulation contained 11,721 atoms (1960 protein atoms) for HEWL, 20,618 atoms (6480 protein or cofactor atoms) for P450cam, and 25,893 atoms (3411 protein or cofactor atoms) for COMT. Starting topology and coordinate files in AMBER format are provided in the Supporting Information.

*MM Equilibration.* The P450cam and HEWL structures were equilibrated with classical (MM) molecular dynamics in AMBER. COMT structures were extracted directly from previous molecular dynamics studies[90]. Minimizations were carried out for 1000 steps with the protein restrained followed by 2000 steps of unrestrained minimization. Following minimization, a 10-ps NVT heating step was carried out to raise the system temperature to 300 K using a Langevin thermostat with collision frequency of 1.0 $ps^{-1}$ and a random seed to avoid synchronization artifacts. This step was followed by a 1-ns NPT equilibration using the Berendsen barostat with a pressure relaxation time of 1 ps. Production dynamics were collected for 100 ns for each protein. The SHAKE algorithm[92] was applied to fix all bonds involving hydrogen, permitting a 2-fs timestep to be used for all MD. For the long-range electrostatics, the particle mesh Ewald method was used with a 10-Å electrostatic cutoff. In the case of P450cam, only a restrained minimization was carried out, and harmonic restraints were employed throughout all dynamics to hold the active site heme, camphor, cysteinate ligand in the crystal structure position (see Supporting Information for restraints).

*QM/MM Simulation*. Snapshots from MD production runs were extracted for QM/MM simulation. The periodic box was post-processed using the center of mass utility in PyMOL[93] to generate the largest possible spherical droplet centered around each protein that was circumscribed by the original rectangular prism periodic box. The resulting system was again processed with tleap to generate a system with spherical cap boundary conditions that was enforced with a restraining potential of 1.5 kcal/mol·$Å^2$. All QM/MM simulations were carried out using TeraChem[31, 94] for the QM portion and AMBER 14[79] for the MM component. The QM region is modeled with density functional theory (DFT) using the range-separated exchange-correlation functional ωPBEh[95] (ω=0.2 bohr$^{-1}$) with the 6-31G*[85] basis set. Select comparisons to results from the B3LYP[96-98] global hybrid functional for FSA were also carried out.

*Partial Charges*. The CSA and FSA schemes rely on evaluation of partial charges obtained from the Voronoi deformation density (VDD) method[70]. The VDD partial charges use a promolecule definition to partition the real space density and are therefore relatively basis-set insensitive. The basis set sensitivity in CSA calculations was carried out on several other basis sets, namely: 6-31G[99], 6-311G, 6-31++G[100], 6-31G**[85], and 6-311++G*, as outlined in the Results and Discussion. FSA comparisons were carried out with the 6-31G*, 6-31G**, and 6-311++G* basis sets.

*CSA and FSA Analysis Methods*. Preparation, automation, and analysis was carried out using in-house python scripts. A tutorial example of the workflow and accompanying scripts for protein analysis are provided on our website (http://hjkgrp.mit.edu/csafsa).

## 4. Results and Discussion

### 4a. Systematic QM Region Determination with CSA and FSA on COMT

*Results of CSA on COMT.* The CSA method was recently introduced[41] as an alternative strategy to systematic radial QM/MM convergence of COMT properties. Briefly, CSA was carried out averaged over 20 snapshots along a pre-computed large-QM/MM methyl transfer reaction barrier. The evaluation of CSA results across the reaction coordinate is motivated by the expected electronic structure variations from charged reactants (S-Adenosyl-L-methionine, SAM and catecholate, CAT) to neutral products (adohomocysteine and O-methylatedcatechol). Similar arguments have been made for choosing intermediate energies to compare in the CDA approach[39]. After averaging, 11 residues have $\Delta q_{\text{RES,av}}$ above a suggested threshold of |0.05 e| (Figure 4). Some of the residues with the strongest charge shifts are expected, e.g.: i) $Mg^{2+}$ coordination sphere residues D141, D169, and N170, ii) catecholate-hydrogen-bonding residues E199, and iii) SAM-hydrogen-bonding residues E64, S72, E90, and H142. Of the four remaining residues, S72 is polar and proximal to SAM, but the detection of the residues V42, A67, and A73 is surprising as they are both nonpolar and not the most proximal to the center of the SAM-catecholate reacting core.

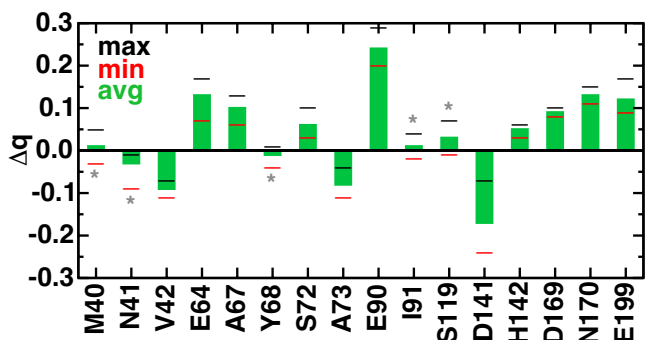

**Figure 4.** CSA analysis holo-apo charge difference (in e) averages (green bars), maximum values (black dashes), and minimum values (red dashes) obtained over 20 snapshots along the methyltransfer reaction coordinate of COMT. Residues selected for maximum or minimum values above |0.05 e| threshold are indicated by a gray asterisk.

A comparison of not just averages but maximum and minimum charge shift values reveals significant ranges in $\Delta q_{\mathrm{RES,av}}$ for several residues, which motivates the identification of 5 additional residues, M40, N41, Y68, I91, and S119, that exceeded the $|0.05\ e|$ threshold for at least one snapshot. We also note the max-min ranges of $\Delta q_{\mathrm{RES}}$ are largest for charged residues, but the averaged contribution is always above threshold for these residues. The lower average contribution for the five polar residues is due to variable interaction across the reaction coordinate combined with lower $\Delta q_{\mathrm{RES,av}}$. For example, M40 and N41 have larger charge shifts only after the transition state. We also note that this method does not detect some proximal residues (i.e., G66 and Y71) that might be selected based on radial distance considerations alone. Previous work demonstrated[41] chemical accuracy could be obtained with respect to the large QM/MM limit for the methyl transfer reaction barrier and enthalpy, particularly when the CSA region containing both the first 11 residues and 5 additional residues is selected. We will return to validation strategies for assessing the CSA results and further testing the relevance of the most unexpected (i.e., nonpolar) residues in Sec. 4d.

*Results of FSA on COMT*. Based on observations of variation in charge shifts across the reaction coordinate, we carried out the FSA method on COMT on three structures: the reactant (R), transition state (TS), and product (P). We compute the electrophilic ($f^-$) and nucleophilic ($f^+$) condensed Fukui functions summed separately over SAM, catecholate, and $Mg^{2+}$, but, as expected, $f^+$ is always zero for catecholate and $f^-$ is always zero for $Mg^{2+}$ (Supporting Information Tables S4-S6). Thus, we focus on shifts of the condensed Fukui functions from median values, i.e., $\Delta f^+$ and $\Delta f^-$ on SAM, $\Delta f^-$ on CAT, and $\Delta f^+$ on $Mg^{2+}$ (Figure 5 and Supporting Information Tables S4-S6). Overall, we designate detected residues as those with an RSS difference obtained over the four Fukui metrics above $|0.05\ e|$ with respect to median values for that structure. From this FSA analysis on three structures, we identify 16 residues that have significant Fukui shifts (FS) in the reactant or transition state and 14 in the product (all shown in Figure 5). This analysis enables identification of the most relevant QM residues: i) $Mg^{2+}$ proximal residues (D169, N170, and D141) are detected more in the product state $\Delta f^+$ of $Mg^{2+}$ than any other component FS contribution, and ii) residues surrounding the reacting species such as M40, N41 and V42 are detected most strongly in the SAM TS $\Delta f$ or catecholate reactant $\Delta f$.

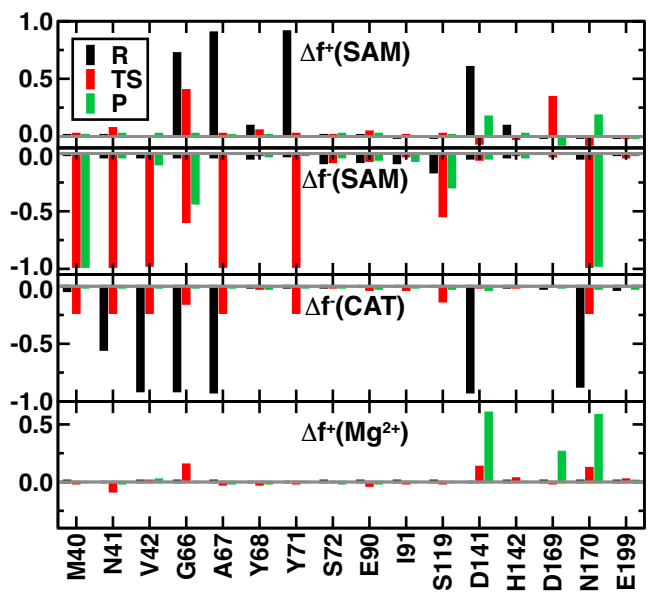

**Figure 5.** FSA analysis of summed condensed Fukui function differences from median for SAM ($\Delta f^+$, top and $\Delta f$, middle top), catecholate (CAT, $\Delta f$, middle bottom), and $Mg^{2+}$ ($\Delta f^+$, bottom) in COMT. All residues above $|0.05$ e$|$ threshold for overall RSS $\Delta f$ in the reactant (R, black bar), transition state (TS, red bar), or product (P, green bar) geometry are shown with single letter residue name and number. Only A67 and Y71 are below this threshold for the products. The range spanned in all graphs is 1.1 e.

Overall comparison of the CSA and FSA results on COMT reveals the two approaches to be in very good agreement (Figure 6). In total, both approaches detect 14 of the same residues, with FSA not detecting the i) E64 residue that hydrogen bonds to SAM and S72, and ii) A73, one of the nonpolar residues identified by CSA. Conversely, two proximal residues to the substrate core, G66 and Y71, were not detected by CSA but are identified by FSA. Given the distinct nature of the two approaches, with FSA being carried out on core substrates plus one additional residue, whereas CSA is carried out in large QM regions, this agreement becomes even more impressive. Analyzing the FSA results for the two residues detected only with this method (i.e., G66 and Y71) reveals that Y71 is one of the two residues detected only in the reactant and transition state with FSA (Supporting Information Table S6). The sole detection of G66 by FSA could potentially be rationalized as a false positive due to boundary effects. Within FSA, each residue is added as a single hydrogen terminated residue, but for a residue as small as glycine, this termination may alter the local electronic environment. We will reconsider this possibility in validating the two methods (see Sec. 4d).

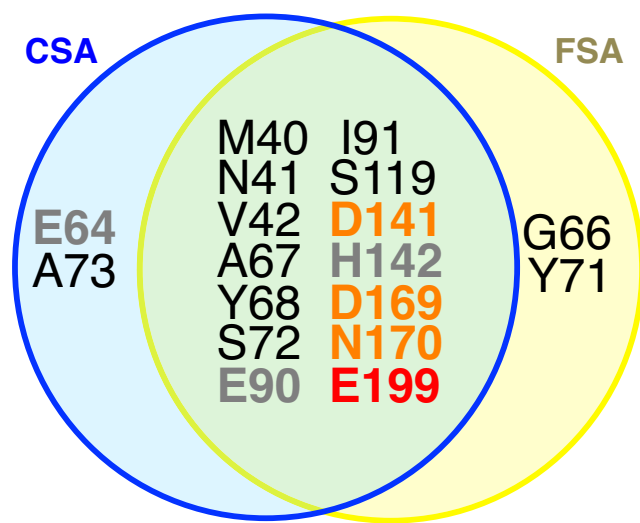

**Figure 6.** Venn diagram of residues indicated by single letter code and number obtained with CSA (left, blue circle) and FSA (right, yellow circle) for COMT. $Mg^{2+}$ coordination sphere residues are annotated in bold orange, and residues that hydrogen bond SAM and catecholate are annotated in bold gray and bold red, respectively.

We also carried out a comparison to CDA, wherein a strong effect of MM point charge deletion on the relative reaction energies for COMT methyl transfer is used to identify that a residue belongs in the QM region (see Supporting Information Table S7). Using a QM region that consists of the core reacting substrates for the CDA analysis, we find high sensitivity to strongly charged residues (e.g., aspartate, glutamate, or lysine) that limits the robustness of CDA for determining optimal QM regions in COMT compared to the FSA or CSA methods. Residues previously identified to contribute strongly to the methyl transfer reaction mechanism[41] were not detected in this approach (e.g., V42, E90), whereas remote charged residues were detected (e.g., E34, E27, K5). CDA may be more suitable for augmenting larger QM regions after construction by chemical intuition, as has been previously suggested[39].

## 4b. Evaluating CSA/FSA Basis Set and Method Sensitivity

In order to determine the best practices for applying the CSA and FSA methods introduced, we now consider the sensitivity of the predicted QM regions for COMT with these methods to both the basis set and electronic structure method with which the analysis is applied. For CSA and FSA, we monitor the electron density in QM/MM environments through partial charge descriptors. Partial charges are well-known[70] to be sensitive to the partitioning scheme and basis set employed. However, CSA and FSA rely on charge differences, which should limit basis set sensitivity, and we have combined these approaches with a real-space partitioned charge density approach that has been demonstrated[70] to be relatively robust. With regard to electronic structure method choice, it has been shown[101] that density delocalization errors are increasingly common as gaps become small (e.g., in semi-local DFT methods), and we[41] and others[102-103] have demonstrated gap closing in large biomolecular systems with semi-local DFT.

The CSA method requires single point calculations with and without core substrates, which we have computed in 940 atom (968 with link atoms, 56 protein residues) radial QM regions on COMT, following previous QM region convergence studies[41]. By construction, residues detected with CSA must be included in this initial QM region, and a very small QM region might lead to increased boundary effects on the CSA results. In order to study larger basis sets (e.g., 6-311++G*) we also carried out CSA using a 534 atom (560 with link atoms, 29 protein residues) radial QM region. The QM region obtained from CSA is determined from the residues that have > |0.05 e| shift from the holo to apo model of the enzyme. Regardless of basis set considered, i.e., 6-31G, 6-31G*, 6-31G**, 6-311G, 6-31++G, or 6-311++G*, the eleven residues (V42, E64, A67, S72, A73, E90, D141, H142, D169, N170, and E199) identified through structurally-averaged CSA with ωPBEh/6-31G in previous work[41] were also detected in the 534-atom QM/MM CSA calculation (Supporting Information Table S8).

Quantitatively, average maximum variations across residues between basis sets were around 0.05 e, although some residues exhibited differences in the charge shift as large as 0.22 e across methods (Supporting Information Table S8). In addition to the core 11 residues identified in previous work, some basis sets separately detected M40 (6-31G, 6-311G, 6-311++G*), I91 (6-211G, 6-311++G*), or S119 (6-31G*, 6-31G**), which were part of 5 residues previously added to the original 11 after identifying residues that had large charge shifts for only part of the reaction coordinate. Comparisons of basis set dependence on the larger 940-atom QM region with 6-31G, 6-31G*, 6-311G, and 6-31G** basis sets shows reduced basis set sensitivity with respect to the smaller QM region, with all basis sets predicting the 11 residues identified in initial studies and only one (6-311G) also detecting M40 (Supporting Information Table S9). Overall, the reliance on qualitative detection of charge shifts through summed partial charges greatly mitigates basis set sensitivity of CSA.

Given the limited variation observed in partial charge differences in CSA and the reliance on relatively small QM regions in FSA of the core substrates plus one additional residue, we directly compared the primary basis set (6-31G*) with two larger basis sets studied in this work (6-31G** and 6-311++G*) on COMT. In both cases, we selected the median distant Fukui result on single residues as the reference for computing RSS Fukui differences, and carried out all calculations with the range-separated hybrid, ωPBEh[95]. Comparison of the qualitative identification of FSA residues in the three basis sets reveals that the same core residues are identified, despite significant changes in the reference median condensed Fukui function values with the larger basis sets (see Supporting Information Tables S10-S11). Larger basis sets may warrant tuning of the range-separation parameter to higher values than were employed here (ω=0.2 bohr$^{-1}$) in order to ensure any unphysical charge transfer is eliminated and that molecular orbital energies, which are probed by the FSA approach, are accurately represented.

Another concern, particularly with the application of the FSA method, is the electronic structure method employed. Asymptotically correct exchange, already motivated in previous work on biomolecules[28, 103], is essential for proper prediction of charge separation across large distances in the QM/MM simulations. Furthermore, well-known delocalization errors[104] in semi-local functionals and global hybrids can produce too small gaps and incorrect molecular orbital energetics, impacting the calculation of the condensed Fukui function in the FSA approach. In fact, we observe numerous false positives for FSA when carried out with the B3LYP functional in comparison to ωPBEh (see nonzero values for B3LYP on zero value of x-axis for ωPBEh in Figure 7 and Supporting Information Table S12). Many of these false positives are likely a result of poor descriptions by B3LYP of non-contiguous QM regions. The functional choice is less critical for CSA, as partial charges on the neutral system are a much less stringent test for an exchange-correlation functional than the redistribution of charge with electron addition or removal (see Supporting Information Table S13). Overall, we recommend a range-separated hybrid (e.g., ωPBEh) for either method with a polarized double-zeta basis set, which will allow for the same method and basis set combination in CSA, FSA, and production dynamics or geometry optimizations.

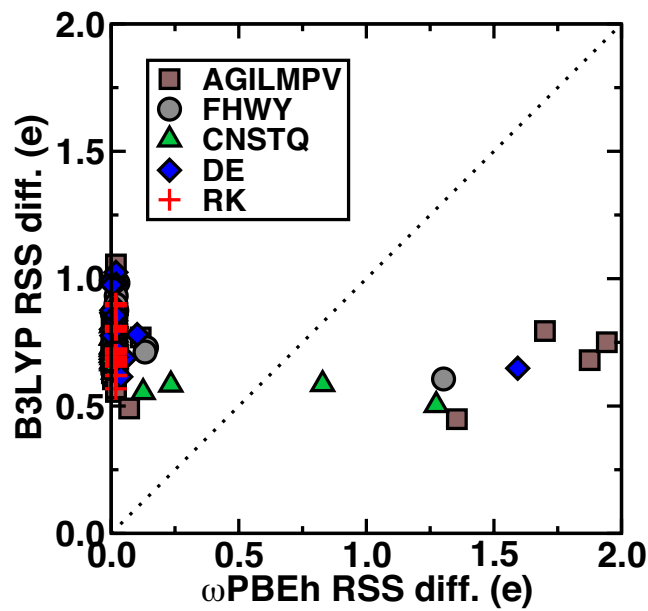

**Figure 7.** Parity plot of B3LYP/6-31G* RSS differences in condensed Fukui functions for residues compared to ωPBEh/6-31G* RSS differences in condensed Fukui functions. Each residue is grouped by class: hydrophobic (brown squares), aromatic (grey circles), polar (green triangles), negatively charged (blue diamonds), and positively charged (red crosses), and the members of each group are indicated by their single letter residue code. A y = x line is indicated with black dashes.

## 4c. Structural Averaging and Sensitivity

Structural averaging was previously motivated[41] for CSA by averaging over up to 20 snapshots across the methyltransfer coordinate. However, it may be desirable to apply CSA and FSA to starting structures before minimum energy pathways have been evaluated with detailed QM/MM study. We thus consider the sensitivity of the two methods to the geometric structure upon which they are applied. In addition to the 20 structures previously sampled, we also compared CSA results on representative MM-equilibrated monodentate and bidentate catecholate structures[90], the unequilibrated crystal structure[74], and the QM-optimized reactant structure studied previously[41] (Supporting Information Table S14). Overall, we observe equivalent CSA detection of 11 key residues for the MM-equilibrated and QM-optimized structures. The crystal structure differs from the other three with a K144 proximal to the catecholate, which CSA detects for this structure only. The initial protonation state of the crystal structure also does not preserve the E64-S72 or E90-SAM hydrogen bonds observed in all other structures, leading to no detection of E90 or S72 for the crystal structure (Supporting Information Figure S1). The relative proximity of M40 to the substrates in the crystal leads to its detection in the reactant crystal structure, whereas this residue was previously detected only in snapshots obtained later in the reaction coordinate.

Therefore, we recommend at least preliminary equilibration with MM to generate snapshots, but no computational pre-optimization with QM is required for CSA. This result is reassuring of the robustness of CSA because the MM structures have been established in COMT to be quite distinct from those obtained with large-QM region QM/MM optimization[41, 47, 90]. FSA demonstrates even further reduced structural dependence than the CSA approach, with the same 16 residues detected for both QM reactant and transition state structures, and only two residues not detected (i.e., G66 and Y71) if the product is taken as the sole reference (see Figure 5 and Supporting Information Tables S4-S6).

**4d. Validation Method for Verifying Systematically Detected QM Regions**

CSA and FSA are sensitive to the thresholds applied for detection, which, if too conservative, could lead to false positives. Earlier work demonstrated[41] that CSA-derived QM regions produced excellent agreement with larger QM/MM models for methyl transfer energetics. In order to broadly test the effect of CSA- or FSA-detected residues on large QM region construction, we introduce two density-based validation metrics, the utility of which we verify with methyltransfer reaction energetics. In both cases, we start by constructing a QM region consisting of all residues detected by the CSA/FSA methods and compute the summed-over-residue partial charges of each core active site residue as well as the surrounding protein residues. Next, we leave out each non-core residue from the QM region, treating it instead with MM, and recompute by-residue partial charge sums. The first metric that indicates the importance of the omitted residue is the RSS charge difference for the core active site residues (ASR) due to moving the residue from QM to MM regions (ASR shift, or ASRS):

$$\text{ASRS(RES)} = \sqrt{\sum_j^{\text{ASR}} (q_j^{\text{RES} \in \text{QM}} - q_j^{\text{RES} \in \text{MM}})^2} \qquad (11)$$

Thus this metric indicates the impact of a specific QM residue on the electronic environment and charge distribution of the active site residues. In order to probe higher-order effects where these residues interact with each other in addition to the core active site residues, we also compute the root mean squared (RMS) difference in partial-charge sums on all residues besides the active site residue (background residues or BGD), that have large values (RES') when the QM→MM transformation is carried out for a given residue:

$$\text{BGD(RES)} = \sqrt{\frac{1}{N} \sum_k^{\text{RES'} \notin \text{ASR}} (q_k^{\text{RES} \in \text{QM}} - q_k^{\text{RES} \in \text{MM}})^2} \quad . \qquad (12)$$

We anticipate this second metric to outweigh the first metric if the residue is contributing only indirectly to the electronic environment at the active site. We also expect that residues detected by CSA but not by FSA may have larger BGD(RES), owing to the many-body effects that may be detected by CSA but not by FSA. Comparison of the two quantities reveals that the direct charge metric dominates and is significant for the majority of residues identified with CSA or FSA (Figure 8, see also Supporting Information Table S15). However, in several notable exceptions, Y71, N41, Y68, and E64, the indirect charge difference outweighs the direct substrate difference.

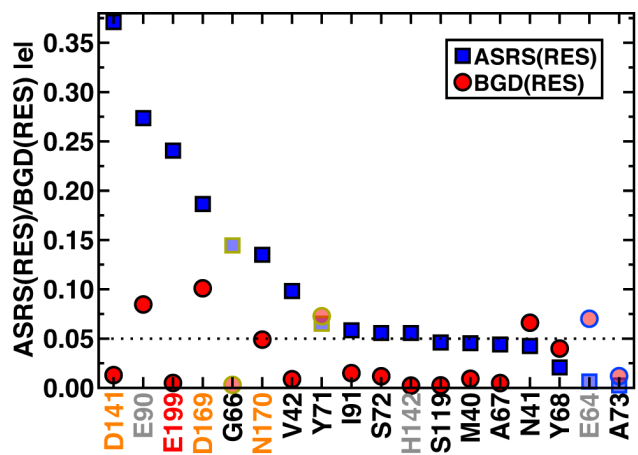

**Figure 8.** ASRS(RES) (blue squares) and BGD(RES) (red circles) partial charge validation metrics in |e| sorted by decreasing ASRS(RES) for the 18 residues identified with CSA or FSA. A 0.05 |e| cutoff is indicated by a black dashed line. The CSA-only residues (E64 and A73) are partially filled and outlined in blue, and the FSA-only residues (G66 and Y71) are partially filled and outlined in yellow. $Mg^{2+}$ coordination sphere residues are colored in orange, and residues that hydrogen bond SAM and catecholate are colored in gray and red, respectively.

The charge density cross-validation analysis suggests several QM regions that can be constructed on the basis of leaving out those residues (e.g., E64, Y68, A73) that have a limited effect on the direct density metrics. We generated several QM models ranging from aggressively pruned to less pruned based on the validation metrics: the top 7, 11, and 15 residues obtained through the RSS metric produced new models **R7**, **R11**, and **R15**, respectively (Figure 9). We also constructed a 17-residue model in which we left out one of the bottom three residues from the direct RSS metric: **R17-1** (E64 omitted), **R17-2** (Y68 omitted), and **R17-3** (A73 omitted), as indicated in Figure 9.

| | M40 | N41 | V42 | E64 | G66 | A67 | Y68 | Y71 | S72 | A73 | E90 | I91 | S119 | D141 | H142 | D169 | N170 | E199 |
|---|---|---|---|---|---|---|---|---|---|---|---|---|---|---|---|---|---|---|
| CSA(16) | ■ | ■ | ■ | ■ | □ | ■ | ■ | □ | ■ | ■ | ■ | ■ | ■ | ■ | ■ | ■ | ■ | ■ |
| FSA(16) | ■ | ■ | □ | ■ | ■ | ■ | ■ | ■ | ■ | □ | ■ | ■ | ■ | ■ | ■ | ■ | ■ | ■ |
| R7 | □ | □ | ■ | ■ | ■ | □ | □ | ■ | □ | ■ | □ | ■ | □ | □ | ■ | □ | ■ | ■ |
| R11 | □ | □ | ■ | ■ | □ | ■ | ■ | □ | ■ | ■ | ■ | ■ | □ | □ | ■ | ■ | ■ | ■ |
| R15 | ■ | ■ | ■ | ■ | ■ | ■ | ■ | ■ | ■ | ■ | ■ | ■ | ■ | ■ | ■ | ■ | ■ | ■ |
| R17-1 | ■ | ■ | ■ | □ | ■ | ■ | ■ | ■ | ■ | ■ | ■ | ■ | ■ | ■ | ■ | ■ | ■ | ■ |
| R17-2 | ■ | ■ | ■ | ■ | ■ | ■ | □ | ■ | ■ | ■ | ■ | ■ | ■ | ■ | ■ | ■ | ■ | ■ |
| R17-3 | ■ | ■ | ■ | ■ | ■ | ■ | ■ | ■ | ■ | □ | ■ | ■ | ■ | ■ | ■ | ■ | ■ | ■ |

**Figure 9.** Definition of QM region models: CSA 16 residue model, FSA 16 residue model, top 7, 11, and 15 residues from validation analysis, and 17 residue models 1, 2, and 3 obtained by leaving out one of the last three residues from the 18 residue model. The protein residues included in the QM region are indicated with a gray shaded square.

In all cases, we compare the computed activation energies, $E_a$, and reaction enthalpies, $\Delta E_{rxn}$, to results obtained with a nearly 1000-atom QM region in QM/MM calculations through an RSS error metric:

$$\mathrm{RSS}(E,\mathrm{M}) = \sqrt{\left(E_a^{\mathrm{M}} - E_a^{\mathrm{ref}}\right)^2 + \left(\Delta E_{rxn}^{\mathrm{M}} - \Delta E_{rxn}^{\mathrm{ref}}\right)^2} \qquad (13)$$

With respect to the large QM region reference, RSS energetic errors range from 0.3 to 3.0 kcal/mol (Table 1). The range of errors is derived most from significant differences in the reaction enthalpy rather than the activation energy where the regions span a 1.2 kcal/mol range from 14.9 to 16.1 kcal/mol. The QM regions with the largest errors at 3.0 and 2.2 kcal/mol, respectively, are the two most aggressively pruned, **R7** and **R11** at 156 and 224 atoms.

**Table 1.** Comparison of the activation energy ($E_a$), reaction enthalpy ($\Delta E_{rxn}$), and RSS error with respect to a large, 56-residue QM/MM simulation all in kcal/mol along with number of atoms (protein and cofactor atoms) for models with model definitions as detailed in Figure 9.

| Model | no. atoms | $E_a$ (kcal/mol) | $\Delta E_{Rxn}$ (kcal/mol) | RSS Error (kcal/mol) |
|---|---|---|---|---|
| CSA (16) | 296 | 16.1 | -10.8 | 0.4 |
| FSA (16) | 299 | 15.6 | -11.2 | 0.3 |
| R7 | 156 | 14.9 | -14.0 | 3.0 |
| R11 | 224 | 15.7 | -9.0 | 2.2 |
| R15 | 278 | 15.8 | -12.3 | 1.1 |
| R17-1 | 309 | 15.7 | -12.4 | 1.2 |
| R17-2 | 303 | 15.7 | -11.6 | 0.4 |
| R17-3 | 314 | 15.7 | -11.5 | 0.4 |
| Ref. | 940 | 15.9 | -11.2 | -- |

The lowest observed RSS errors at 0.3-0.4 kcal/mol are for the CSA or FSA models, which both have 16 residues and about 296 atoms. The bulk of the discrepancy in other models comes from either overestimating exothermicity (**R7**) or underestimating it (**R11**). If we are purely concerned with reproducing an activation energy, then **R7** or **R11** may be deemed sufficient, but this and previous analysis reveals that other characteristics, such as the electronic environment and reaction enthalpy will not necessarily be well-described in these smaller models. Overall, these results highlight that slight differences in the CSA- and FSA- constructed QM regions yield small differences but omission of critical residues produces large errors along the reaction coordinate at activation energies or in reaction enthalpies. The CSA and FSA regions perform comparably, and the validation metrics combined with energetic analysis suggest that both low RSS and RMS are necessary to exclude residues from a QM region. Thus, this analysis indicates

A73 represents a likely false positive obtained with CSA and that Y68, although detected by both procedures, can be omitted without loss of accuracy in energetic properties.

A further consideration in choosing between the two methods is in their computational cost. Although large-scale DFT calculations with a moderate basis set are becoming increasingly routine, an attractive feature of FSA is its inherent parallelizability of a smaller QM calculation that contains only core active sites and one additional residue. Comparison of timings obtained on COMT with TeraChem[31, 94] on 1 nVidia GTX 970 for FSA or 4 nVidia GTX 970s for CSA reveals that the CSA approach still requires fewer overall GPU hours (5.6 GPU hours for CSA vs. 11.4 for the cheapest FSA approach). This relative efficiency of CSA is due to the fact that three calculations must be carried out for each FSA residue (Table 2). Nevertheless, the shortest wall time is achieved with FSA at as little as 6 minutes if all calculations are executed in parallel, compared to 45 minutes with CSA. Neither approach on a single snapshot or even averaged over several snapshots amounts to a substantial computational cost compared to the effort required for geometry optimization, transition state search, or dynamic sampling typically carried out during mechanistic enzyme study. The choice of one method over another may come down to the nature of the active site (e.g., with or without covalently bound substrates, see secs. 4e-4f) or the ease of applying the CSA approach if presently available software and hardware prevent simulation of QM regions larger than 500 atoms in size.

**Table 2.** Comparison of the timing of the FSA and CSA approaches on a single snapshot of COMT with ωPBEh/6-31G*.

|  | FSA med. | FSA large | FSA all | CSA med. | CSA large |
|---|---|---|---|---|---|
| res. in calc. | 1 | 1 | 1 | 29 | 56 |
| # of calcs. | 3 | 3 | 3 | 1 | 1 |
| # of res. x calcs. | 29x3 = 87 | 56x3 = 168 | 214x3 = 642 | 29 | 56 |
| # of GPUs | 1 | 1 | 1 | 4 | 4 |
| walltime/calc. | 2.7-**5.8** | 2.7-**5.8** | 2.7-**5.8** | 8.4-**9.2** | 40-**44** |
| (min.) | | | | | |
| GPU hours | 5.9 | 11.4 | 43.7 | 1.2 | 5.6 |

### 4e. Demonstrating CSA and FSA on HEWL

Hen eggwhite lysozyme[105] (PDB ID: 2VB1[72]) is a 129-residue (1960 atom) enzyme that has been previously employed in QM/MM convergence studies by Fuxreiter and coworkers[42]. Using the semi-empirical PM3 method[106], these authors showed that forces converged in the polar region of the enzyme surrounding residue Glu35 only when about 500 atoms were included in the QM region. The identified residues inside that region included adjacent waters, Phe34, Ser36 as well as nearby Asp52, Gln57, and a portion of Asn44[42]. The authors also studied the slow convergence of free energy barriers for proton transfer between the central Glu35 and a neighboring water molecule[42]. Here, we carried out CSA on HEWL, defining the central core residues as Glu35 and an adjacent water molecule, which we mutated to an alanine and removed, respectively, to form the apo HEWL enzyme. Due to the smaller perturbation of mutating Glu to Ala rather than complete removal of residues as in COMT, we reduced the threshold for CSA charge detection to |0.04 e| and carried out analysis on both a reactant structure with glutamate/water and a product structure with glutamic acid/hydroxyl ion.

The small size of HEWL enabled us to place the entire protein and several water molecules inside the QM region for the CSA procedure. The charge shift analysis revealed some variations between results on reactant and product structure, and detection in at least one reactant or product structure led to identification of 15 residues overall (Figure 10). Several water molecules adjacent to the Glu35 were also identified as critical (Supporting Information Tables S16-S17). All residues identified previously[42] as essential for minimizing force errors were identified by CSA except for Asn44. In total, the CSA-derived QM/MM model contains 297 QM atoms (244 protein residue atoms, 20 link atoms, and 11 water molecules), which is reduced from the 500-atom-sized radial regions suggested in the original work by Fuxreiter and coworkers[42]. An alternative prescription for CSA in which Glu35 is removed completely with link-atom termination of Phe34 and Ser36 produces similar qualitative results, albeit at the cost of disrupting the backbone of the adjacent residues and overestimating their charge shifts (see Supporting Information Table S18).

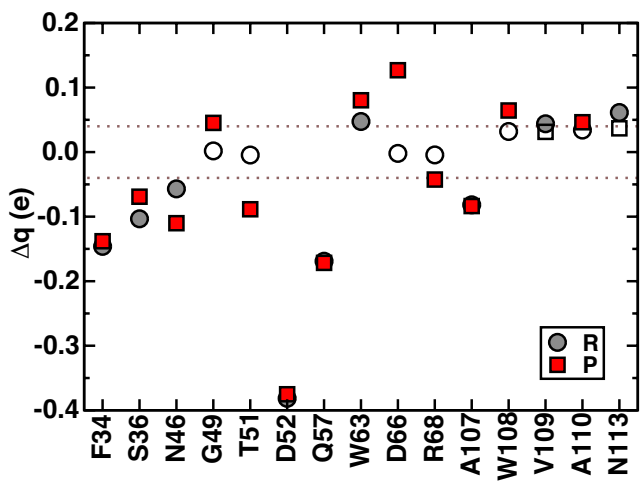

**Figure 10.** CSA apo-holo charge difference (in e) for HEWL reactant (gray circles) and product structures (red squares). The minimum threshold values |0.04 e| are shown as brown dashed lines, and individual reactant or product snapshots that fall below this value are shown as empty symbols.

Upon carrying out FSA on a reduced subset of residues within 9 Å of the carboxylate O⁻, residues detected in either the reactant or product state are comparable to those obtained from CSA (Figure 11 and Supporting Information Table S19). The main differences between the CSA and FSA results include detection of the Asn44 residue identified in previous work as well as Lys33, which leads to a QM region that contains 34 additional protein residue atoms.

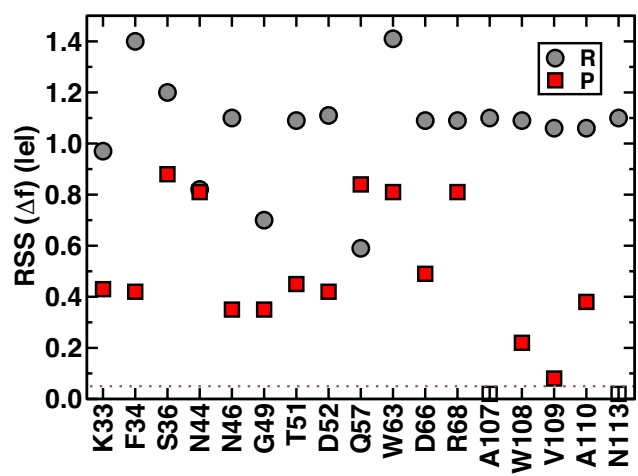

**Figure 11.** FSA RSS (in e) for HEWL reactant (gray circles) and product structures (red squares). The minimum threshold values |0.05 e| are shown as brown dashed lines, and individual reactant or product snapshots that fall below this value are shown as empty symbols.

We validate the CSA- and FSA-detected QM regions by computing the enthalpy of Glu35 protonation relative to a minimal QM region (min) consisting of only Glu35 and the water molecule that is deprotonated as well as a 500-atom QM region (large) that is obtained from any residue within 9 Å of the central Glu35 $O^-$ to be consistent with previous work[42] (a residue list is provided in the Supporting Information). Previous estimates of the free energy of deprotonation with semi-empirical QM[42] work had indicated a reduction from 23 kcal/mol to 19 kcal/mol moving from a minimal model to one that included atoms up to 6 Å from the central Glu35 $O^-$. Here we observe similar variations in protonation enthalpies with increasing QM region size but with distinct absolute QM enthalpies compared to the previously reported semi-empirical free energies (Table 3). Overall, the CSA and FSA QM regions are within chemical accuracy of the reaction enthalpy produced with the larger QM region, whereas there is a 5-kcal/mol overestimation of the reaction enthalpy with the small QM region, consistent with previous ranges of free energies. These results on HEWL suggest that the methyl capping approach for active site modification is a suitable strategy for QM region detection with covalently bound residues.

**Table 3.** Proton transfer energetics for HEWL with minimal, large radial, CSA, and FSA QM regions.

| Model | $\Delta E_{rxn}$ (kcal/mol) |
|-------|-------------|
| min   | 29.8 |
| large | 24.4 |
| CSA   | 24.9 |
| FSA   | 24.8 |

## 4f. Demonstrating CSA and FSA on P450cam

Cytochrome P450cam (PDB ID: 1DZ9[73]) has been previously employed in a study of differences between QM properties in QM/MM calculations with mechanical-, electronic-, and polarizable-embedding[23]. The QM treatment of the active site consisted of an iron-oxo heme with coordinating Cys357 and camphor substrate. Hirao and coworkers[23] found that reduction of charges along the protein backbone adjacent to Cys357, namely of Leu356 and Leu358, was helpful in reducing overpolarization of Cys357 in the QM region, which had prevented observation of spin density on the porphyrin. The authors also demonstrated that polarizable-embedding did not appreciably change the predictions of spin-polarization or estimations of QM-MM interaction energies[23]. In the active site of P450cam, the heme hydrogen bonds to Arg299, Arg112, and His355, and camphor is held in place by a hydrogen bond to Tyr96.

Here, we carried out CSA on a 56-residue and 10-water QM region in a QM/MM model of P450cam, defining the core region as heme, Cys357, and the camphor substrate. The heme and camphor were removed, and Cys357 was mutated to an alanine. Analysis on a reactant model of P450cam was carried out, and we used the |0.04 e| threshold again due to reduced perturbation of the Cys357 to Ala mutation. In total, 9 protein residues were detected along with

one water from CSA analysis (Figure 12 and See Supporting Information Table S20). Detected residues included the Arg112 and Arg299 hydrogen bond partners to the heme as well as several residues adjacent to camphor (Leu244 and Asp297) or to the cysteinate residue that coordinates heme (Leu358, Gly359, and Gln360). The constructed QM region of 9 residues, water, heme, camphor, and cysteinate consists of 271 atoms.

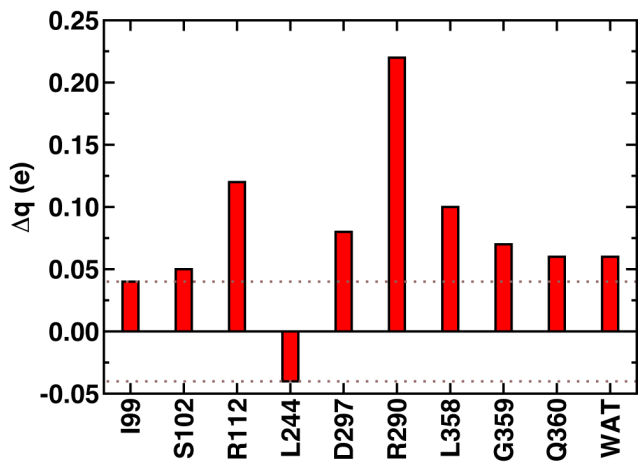

**Figure 12.** CSA apo-holo charge difference (in e) for P450cam reactant structure. The minimum threshold values |0.04 e| are shown as brown dashed lines.

We also carried out FSA analysis on P450cam, where we employ a doublet spin state for the neutral reference system and add or remove electrons from majority spin in both cases to produce singlets (Supporting Information Table S21). The FSA analysis identifies 12 residues and one water molecule, including most of those obtained with CSA (Figure 13). Residues detected by FSA but not CSA include the heme hydrogen bonding His355 and the camphor hydrogen-bonding partner Tyr96 as well as heme-adjacent Thr101. The constructed QM region of 12 residues, water, heme, camphor, and cysteinate consists of 331 atoms. The Ile99 and Ser102 residues detected by CSA were not detected by FSA, although we note that structural averaging as was carried out in COMT could bring the two methods into increased agreement.

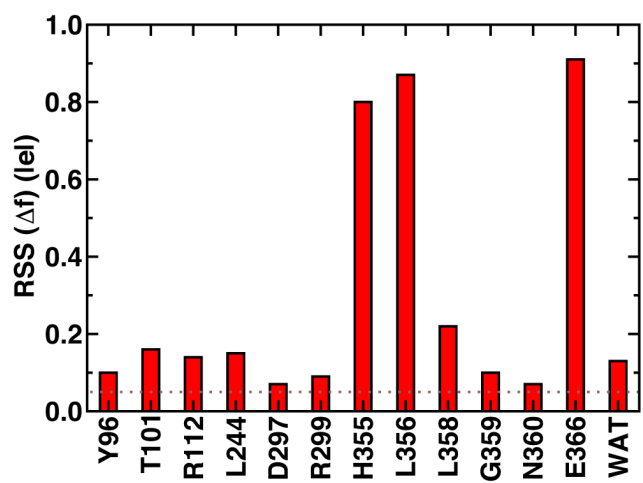

**Figure 13.** FSA RSS (in e) for P450cam reactant structures. The minimum threshold value |0.05 e| is shown as a brown dashed line.

Both CSA and FSA QM regions demonstrate consistency with large QM/MM simulations in terms of partial charge distribution on the cysteinate in the active site. Hirao and coworkers[23] previously noted unphysical spin-polarization on the cysteinate that was diminished if boundary MM charges were reduced, but analysis of changes in total charge on the cysteinate with increasing QM region was not carried out. Comparison of cysteinate-summed VDD charges for the small (cysteinate, heme, and camphor), large QM (56-residue region as defined for CSA), CSA, and FSA regions reveals greater overall net negative charge on the cysteinate for the minimal region (Table 4). The CSA, FSA, and large QM regions quantitatively agree, whereas the minimal QM model carries a larger negative sulfur partial charge and by-residue cysteinate partial charge. Importantly, this change in net partial charge cannot be addressed by adjusting boundary charges alone, although boundary charge adjustment could be carried out in concert with CSA/FSA-constructed QM regions.

**Table 4.** Partial charge sums on cysteinate and partial charge on cysteinate sulfur atom (in e) for P450cam with small, large, CSA, and FSA QM regions.

| Model | Cys357 $q$ | S $q$ |
| --- | --- | --- |

|       | (e)   | (e)   |
|-------|-------|-------|
| small | -0.42 | -0.53 |
| large | 0.10  | -0.44 |
| CSA   | 0.09  | -0.44 |
| FSA   | 0.09  | -0.45 |

## 5. Conclusions

We have developed and extended two complementary methods for systematic determination of QM regions in QM/MM simulations. First, in charge shift analysis, the electronic reorganization of the protein environment around the core active site is probed through identification of how by-residue partial charges shift when the core substrates are removed completely or through removal and methyl capping of covalently bound substrates. Although we had previously suggested[41] the promise of CSA for non-covalently bound substrates, we have now demonstrated its broad applicability across a range of active sites. Secondly, in Fukui shift analysis, the interaction between a residue and the core substrates is probed via changes in the condensed Fukui functions in the protein residues. We demonstrated that both methods produce atom-economical QM regions ca. 15 residues or 290 atoms in size in very good agreement on a model enzyme, COMT, substantially reduced with respect to radially converged QM regions ca. 500-600 atoms in size or more. The benefit of these analysis methods is two-fold: i) first, we identify key residues that interact with the core active site rather than simply obtaining information about a general distance range as in radial convergence studies, and ii) we obtain QM regions that become tractable for extensive sampling to obtain free energies or numerous transition states. The computational cost for both analysis methods was revealed to be modest, but FSA remains more suitable for cases where highly-parallelizable calculations are favored or large-QM calculations (ca. 500-1000 atoms) are not feasible with the user's available hardware and software. We demonstrated that both methods have limited sensitivity to structure choice and basis set size, but we motivated the use of asymptotically-correct, range-separated hybrid functionals, particularly for the FSA method.

We introduced a validation strategy for the CSA and FSA methods to judge any false positives that may arise. In the validation strategy, we generated a QM model with all detected QM region residues and returned individual residues from the QM region back to the MM region, identifying how i) core substrate partial charges shifted and ii) how other key residue charges shifted. The former was a direct measure of a residue's interaction with the substrates, whereas the latter measured indirect effects through residue-residue interactions. The majority of residues detected through CSA or FSA were shown to be important, as confirmed by both large effects on charge density when the residue is treated with MM instead of QM and from calculated reaction energetics. The validation strategy provided some further atom-economy by identifying the residues most inconsequentially omitted from QM regions, although CSA or FSA regions may be employed without further reduction and validation with no reduction in accuracy.

Finally, we demonstrated the broad utility of our methods on the proton transfer reaction energetics in hen eggwhite lysozyme and the partial charges on heme-coordinating residues in the active site of P450cam. For these additional proteins, covalently bound active site residues necessitated the demonstration of methyl capping in CSA. Despite this additional challenge, CSA and FSA were useful for identifying quantitatively converged QM regions ca. 200-300 atoms in size, smaller than previous observations in the case of HEWL that suggested 500 atoms were necessary to obtain converged active site properties. Overall, both the new methods for systematic determination of QM regions and validation tools introduced are expected to provide broad utility for improving the robustness of multi-scale modeling efforts.

## ASSOCIATED CONTENT

**Supporting Information**. Topology and coordinate files for amber simulations; restraints for P450cam simulation; protonation states of residues in COMT, P450cam, and HEWL; FSA results on COMT reactant, transition state, and product; CDA results on COMT reaction energetics; CSA basis set sensitivity on COMT in moderate and large model; FSA basis set sensitivity on COMT; B3LYP FSA results on COMT; B3LYP CSA results on COMT; CSA structural sensitivity for COMT; charge difference validation metrics; CSA and FSA results on P450cam; CSA and FSA results on HEWL; CSA results on HEWL with complete residue removal.

## AUTHOR INFORMATION

**Corresponding Author**

*email: hjkulik@mit.edu phone: 617-253-4584

**Notes**

The authors declare no competing financial interest.

## ACKNOWLEDGMENT

H.J.K. holds a Career Award at the Scientific Interface from the Burroughs Wellcome Fund, which supported this work. This work was carried out in part using computational resources from the Extreme Science and Engineering Discovery Environment (XSEDE), which is supported by National Science Foundation grant number ACI-1053575. The authors thank Adam H. Steeves for providing a critical reading of the manuscript.

REFERENCES

1.      Gao, J.; Ma, S.; Major, D. T.; Nam, K.; Pu, J.; Truhlar, D. G., Mechanisms and Free Energies of Enzymatic Reactions. *Chem. Rev.* **2006,** *106*, 3188-3209.
2.      Field, M. J.; Bash, P. A.; Karplus, M., A Combined Quantum-Mechanical and Molecular Mechanical Potential for Molecular-Dynamics Simulations. *J. Comput. Chem.* **1990,** *11*, 700-733.
3.      Bakowies, D.; Thiel, W., Hybrid Models for Combined Quantum Mechanical and Molecular Mechanical Approaches. *J. Phys. Chem.* **1996,** *100*, 10580-10594.
4.      Mordasini, T. Z.; Thiel, W., Combined Quantum Mechanical and Molecular Mechanical Approaches. *Chimia* **1998,** *52*, 288-291.
5.      Monard, G.; Merz, K. M., Combined Quantum Mechanical/Molecular Mechanical Methodologies Applied to Biomolecular Systems. *Acc. Chem. Res.* **1999,** *32*, 904-911.
6.      Gao, J. L.; Truhlar, D. G., Quantum Mechanical Methods for Enzyme Kinetics. *Annu. Rev. Phys. Chem.* **2002,** *53*, 467-505.
7.      Rosta, E.; Klahn, M.; Warshel, A., Towards Accurate Ab Initio QM/MM Calculations of Free-Energy Profiles of Enzymatic Reactions. *J. Phys. Chem. B* **2006,** *110*, 2934-2941.
8.      Lin, H.; Truhlar, D., QM/MM: What Have We Learned, Where Are We, and Where Do We Go from Here? *Theor. Chem. Acc.* **2007,** *117*, 185-199.
9.      Warshel, A.; Levitt, M., Theoretical Studies of Enzymic Reactions: Dielectric, Electrostatic and Steric Stabilization of the Carbonium Ion in the Reaction of Lysozyme. *J. Mol. Bio.* **1976,** *103*, 227-249.
10.     Senn, H. M.; Thiel, W., QM/MM Methods for Biomolecular Systems. *Angew. Chem. Int. Ed.* **2009,** *48*, 1198-1229.
11.     Vidossich, P.; Florin, G.; Alfonso-Prieto, M.; Derat, E.; Shaik, S.; Rovira, C., On the Role of Water in Peroxidase Catalysis: A Theoretical Investigation of Hrp Compound I Formation. *J. Phys. Chem. B* **2010,** *114*, 5161-5169.
12.     Carloni, P.; Rothlisberger, U.; Parrinello, M., The Role and Perspective of Ab Initio Molecular Dynamics in the Study of Biological Systems. *Acc. Chem. Res.* **2002,** *35*, 455-464.
13.     Eurenius, K. P.; Chatfield, D. C.; Brooks, B. R.; Hodoscek, M., Enzyme Mechanisms with Hybrid Quantum and Molecular Mechanical Potentials. I. Theoretical Considerations. *Int. J. Quantum Chem.* **1996,** *60*, 1189-1200.
14.     Senn, H. M.; Thiel, W., QM/MM Studies of Enzymes. *Curr. Opin. Chem. Biol.* **2007,** *11*, 182-187.
15.     Monari, A.; Rivail, J.-L.; Assfeld, X., Advances in the Local Self-Consistent Field Method for Mixed Quantum Mechanics/Molecular Mechanics Calculations. *Acc. Chem. Res.* **2012,** *46*, 596-603.
16.     Wang, Y.; Gao, J., Projected Hybrid Orbitals: A General QM/MM Method. *J. Phys. Chem. B* **2015,** *119*, 1213-1224.

17.     Slavicek, P.; Martinez, T. J., Multicentered Valence Electron Effective Potentials: A Solution to the Link Atom Problem for Ground and Excited Electronic States. *J. Chem. Phys.* **2006,** *124*, 084107.

18.     Murphy, R. B.; Philipp, D. M.; Friesner, R. A., A Mixed Quantum Mechanics/Molecular Mechanics (QM/MM) Method for Large Scale Modeling of Chemistry in Protein Environments. *J. Comput. Chem.* **2000,** *21*, 1442-1457.

19.     Zhang, Y.; Lee, T.-S.; Yang, W., A Pseudobond Approach to Combining Quantum Mechanical and Molecular Mechanical Methods. *J. Chem. Phys.* **1999,** *110*, 46-54.

20.     DiLabio, G. A.; Hurley, M. M.; Christiansen, P. A., Simple One-Electron Quantum Capping Potentials for Use in Hybrid QM/MM Studies of Biological Molecules. *J. Chem. Phys.* **2002,** *116*, 9578-9584.

21.     von Lilienfeld, O. A.; Tavernelli, I.; Rothlisberger, U.; Sebastiani, D., Variational Optimization of Effective Atom-Centered Potentials for Molecular Properties. *J. Chem. Phys.* **2005,** *122*, 14113.

22.     Wang, B.; Truhlar, D. G., Combined Quantum Mechanical and Molecular Mechanical Methods for Calculating Potential Energy Surfaces: Tuned and Balanced Redistributed Charge Algorithm. *J. Chem. Theo. Comp.* **2010,** *6*, 359-369.

23.     Thellamurege, N. M.; Hirao, H., Effect of Protein Environment within Cytochrome P450cam Evaluated Using a Polarizable-Embedding QM/MM Method. *J. Phys. Chem. B* **2014,** *118*, 2084-2092.

24.     Ponder, J. W.; Wu, C.; Ren, P.; Pande, V. S.; Chodera, J. D.; Schnieders, M. J.; Haque, I.; Mobley, D. L.; Lambrecht, D. S.; DiStasio Jr, R. A., Current Status of the AMOEBA Polarizable Force Field. *J. Phys. Chem. B* **2010,** *114*, 2549-2564.

25.     Halgren, T. A.; Damm, W., Polarizable Force Fields. *Curr. Opin. Struct. Biol.* **2001,** *11*, 236-242.

26.     Ufimtsev, I. S.; Luehr, N.; Martínez, T. J., Charge Transfer and Polarization in Solvated Proteins from Ab Initio Molecular Dynamics. *J. Phys. Chem. Lett.* **2011,** *2*, 1789-1793.

27.     Nadig, G.; Van Zant, L. C.; Dixon, S. L.; Merz, K. M., Charge-Transfer Interactions in Macromolecular Systems: A New View of the Protein/Water Interface. *J. Amer. Chem. Soc.* **1998,** *120*, 5593-5594.

28.     Kulik, H. J.; Luehr, N.; Ufimtsev, I. S.; Martinez, T. J., Ab Initio Quantum Chemistry for Protein Structure. *J. Phys. Chem. B* **2012,** *116*, 12501-12509.

29.     Ufimtsev, I. S.; Martínez, T. J., Quantum Chemistry on Graphical Processing Units. 1. Strategies for Two-Electron Integral Evaluation. *J. Chem. Theory Comput.* **2008,** *4*, 222-231.

30.     Ufimtsev, I. S.; Martínez, T. J., Quantum Chemistry on Graphical Processing Units. 2. Direct Self-Consistent-Field Implementation. *J. Chem. Theory Comput.* **2009,** *5*, 1004-1015.

31.     Ufimtsev, I. S.; Martínez, T. J., Quantum Chemistry on Graphical Processing Units. 3. Analytical Energy Gradients, Geometry Optimization, and First Principles Molecular Dynamics. *J. Chem. Theory Comput.* **2009,** *5*, 2619-2628.

32.     Isborn, C. M.; Luehr, N.; Ufimtsev, I. S.; Martinez, T. J., Excited-State Electronic Structure with Configuration Interaction Singles and Tamm-Dancoff Time-Dependent Density Functional Theory on Graphical Processing Units. *J. Chem. Theory Comput.* **2011,** *7*, 1814-1823.

33.     Ochsenfeld, C.; Kussmann, J.; Lambrecht, D. S., Linear-Scaling Methods in Quantum Chemistry. *Rev. Comput. Chem.* **2007,** *23*, 1.

34.     Eichkorn, K.; Weigend, F.; Treutler, O.; Ahlrichs, R., Auxiliary Basis Sets for Main Row Atoms and Transition Metals and Their Use to Approximate Coulomb Potentials. *Theor. Chem. Acc.* **1997,** *97*, 119-124.

35.     Eichkorn, K.; Treutler, O.; Öhm, H.; Häser, M.; Ahlrichs, R., Auxiliary Basis Sets to Approximate Coulomb Potentials. *Chem. Phys. Lett.* **1995,** *240*, 283-290.

36.     Flaig, D.; Beer, M.; Ochsenfeld, C., Convergence of Electronic Structure with the Size of the QM Region: Example of QM/MM NMR Shieldings. *J. Chem. Theory Comput.* **2012,** *8*, 2260-2271.

37.     Hartman, J. D.; Neubauer, T. J.; Caulkins, B. G.; Mueller, L. J.; Beran, G. J., Converging Nuclear Magnetic Shielding Calculations with Respect to Basis and System Size in Protein Systems. *J. Biomol. NMR* **2015,** *62*, 327-340.

38.     Fox, S. J.; Pittock, C.; Fox, T.; Tautermann, C. S.; Malcolm, N.; Skylaris, C. K., Electrostatic Embedding in Large-Scale First Principles Quantum Mechanical Calculations on Biomolecules. *J. Chem. Phys.* **2011,** *135*, 224107.

39.     Liao, R. Z.; Thiel, W., Convergence in the QM-Only and QM/MM Modeling of Enzymatic Reactions: A Case Study for Acetylene Hydratase. *J. Comput. Chem.* **2013,** *34*, 2389-2397.

40.     Sadeghian, K.; Flaig, D.; Blank, I. D.; Schneider, S.; Strasser, R.; Stathis, D.; Winnacker, M.; Carell, T.; Ochsenfeld, C., Ribose-Protonated DNA Base Excision Repair: A Combined Theoretical and Experimental Study. *Ang. Chem. Int. Ed.* **2014,** *53*, 10044-10048.

41.     Kulik, H. J.; Zhang, J.; Klinman, J. P.; Martinez, T. J., How Large Should the QM Region Be in QM/MM Calculations? The Case of Catechol O-Methyltransferase. *J. Phys. Chem. B* **2016,** *Just accepted*.

42.     Solt, I.; Kulhanek, P.; Simon, I.; Winfield, S.; Payne, M. C.; Csanyi, G.; Fuxreiter, M., Evaluating Boundary Dependent Errors in QM/MM Simulations. *J. Phys. Chem. B* **2009,** *113*, 5728-5735.

43.     Isborn, C. M.; Goetz, A. W.; Clark, M. A.; Walker, R. C.; Martinez, T. J., Electronic Absorption Spectra from Mm and Ab Initio QM/MM Molecular Dynamics: Environmental Effects on the Absorption Spectrum of Photoactive Yellow Protein. *J. Chem. Theory Comput.* **2012,** *8*, 5092-5106.

44.     Vanpoucke, D. E.; Oláh, J.; De Proft, F.; Van Speybroeck, V.; Roos, G., Convergence of Atomic Charges with the Size of the Enzymatic Environment. *J. Chem. Inf. Model.* **2015,** *55*, 564-571.

45.     Harris, T. V.; Szilagyi, R. K., Protein Environmental Effects on Iron-Sulfur Clusters: A Set of Rules for Constructing Computational Models for Inner and Outer Coordination Spheres. *J. Comput. Chem.* **2016,** *37*, 1681-1696.

46.     Axelrod, J.; Tomchick, R., Enzymatic O-Methylation of Epinephrine and Other Catechols. *J. Biol. Chem.* **1958,** *233*, 702-705.

47.     Zhang, J.; Kulik, H. J.; Martinez, T. J.; Klinman, J. P., Mediation of Donor–Acceptor Distance in an Enzymatic Methyl Transfer Reaction. *Proc. Natl. Acad. Sci. U. S. A.* **2015,** *112*, 7954-7959.

48.     Rudberg, E.; Rubensson, E. H.; Salek, P., Kohn-Sham Density Functional Theory Electronic Structure Calculations with Linearly Scaling Computational Time and Memory Usage. *J. Chem. Theo. Comp.* **2011,** *7*, 340-350.

49.     Challacombe, M.; Schwegler, E., Linear Scaling Computation of the Fock Matrix. *J. Chem. Phys.* **1997,** *106*, 5526-5536.

50.     Skylaris, C.-K.; Haynes, P. D.; Mostofi, A. A.; Payne, M. C., Introducing ONETEP: Linear Scaling Density Functional Simulations on Parallel Computers. *J. Chem. Phys.* **2005,** *122*, 084119.

51.     Bowler, D. R.; Miyazaki, T., O(N) Methods in Electronic Structure Calculations. *Rep. Prog. Phys.* **2012,** *75*, 036503.

52.     VandeVondele, J.; Borštnik, U.; Hutter, J., Linear Scaling Self-Consistent Field Calculations with Millions of Atoms in the Condensed Phase. *J. Chem. Theory Comput.* **2012,** *8*, 3565-3573.

53.     Scuseria, G. E., Linear Scaling Density Functional Calculations with Gaussian Orbitals. *J. Phys. Chem. A* **1999,** *103*, 4782-4790.

54.     Guidon, M.; Hutter, J.; VandeVondele, J., Robust Periodic Hartree-Fock Exchange for Large-Scale Simulations Using Gaussian Basis Sets. *J. Chem. Theo. Comp.* **2009,** *5*, 3010-3021.

55.     Sumner, S.; Söderhjelm, P.; Ryde, U., Effect of Geometry Optimizations on QM-Cluster and QM/MM Studies of Reaction Energies in Proteins. *J. Chem. Theory Comput.* **2013,** *9*, 4205-4214.

56.     Liao, R.-Z.; Thiel, W., Comparison of QM-only and QM/MM Models for the Mechanism of Tungsten-Dependent Acetylene Hydratase. *J. Chem. Theory Comput.* **2012,** *8*, 3793-3803.

57.     Bash, P. A.; Field, M. J.; Davenport, R. C.; Petsko, G. A.; Ringe, D.; Karplus, M., Computer Simulation and Analysis of the Reaction Pathway of Triosephosphate Isomerase. *Biochemistry* **1991,** *30*, 5826-5832.

58.     Hu, L.; Eliasson, J.; Heimdal, J.; Ryde, U., Do Quantum Mechanical Energies Calculated for Small Models of Protein-Active Sites Converge?†. *J. Phys. Chem. A* **2009,** *113*, 11793-11800.

59.     Hu, L.; Söderhjelm, P.; Ryde, U., Accurate Reaction Energies in Proteins Obtained by Combining QM/MM and Large QM Calculations. *J. Chem. Theory Comput.* **2012,** *9*, 640-649.

60.     Massova, I.; Kollman, P. A., Combined Molecular Mechanical and Continuum Solvent Approach (MM-PBSA/GBSA) to Predict Ligand Binding. *Perspect. Drug Discovery Des.* **2000,** *18*, 113-135.

61.     Hopffgarten, M. v.; Frenking, G., Energy Decomposition Analysis. *Wiley Interdiscip. Rev.: Comput. Mol. Sci.* **2012,** *2*, 43-62.

62.     Sumowski, C. V.; Ochsenfeld, C., A Convergence Study of QM/MM Isomerization Energies with the Selected Size of the QM Region for Peptidic Systems†. *J. Phys. Chem. A* **2009,** *113*, 11734-11741.

63.     Hu, L.; Söderhjelm, P.; Ryde, U., On the Convergence of QM/MM Energies. *J. Chem. Theory Comput.* **2011,** *7*, 761-777.

64.     Meier, K.; Thiel, W.; van Gunsteren, W. F., On the Effect of a Variation of the Force Field, Spatial Boundary Condition and Size of the QM Region in QM/MM MD Simulations. *J. Comput. Chem.* **2012,** *33*, 363-378.

65.     Wong, K. F.; Watney, J. B.; Hammes-Schiffer, S., Analysis of Electrostatics and Correlated Motions for Hydride Transfer in Dihydrofolate Reductase. *J. Phys. Chem. B* **2004,** *108*, 12231-12241.

66.     Parr, R. G.; Yang, W., Density Functional Approach to the Frontier-Electron Theory of Chemical Reactivity. *J. Amer. Chem. Soc.* **1984,** *106*, 4049-4050.

67.     Yang, W.; Mortier, W. J., The Use of Global and Local Molecular Parameters for the Analysis of the Gas-Phase Basicity of Amines. *J. Amer. Chem. Soc.* **1986,** *108*, 5708-5711.

68.     Faver, J.; Merz Jr, K. M., Utility of the Hard/Soft Acid– Base Principle via the Fukui Function in Biological Systems. *J. Chem. Theory Comput*. **2010**, *6*, 548-559.

69.     Fukushima, K.; Wada, M.; Sakurai, M., An Insight into the General Relationship between the Three Dimensional Structures of Enzymes and Their Electronic Wave Functions: Implication for the Prediction of Functional Sites of Enzymes. *Proteins: Struct., Funct., Bioinf*. **2008**, *71*, 1940-1954.

70.     Guerra, C. F.; Handgraaf, J. W.; Baerends, E. J.; Bickelhaupt, F. M., Voronoi Deformation Density (VDD) Charges: Assessment of the Mulliken, Bader, Hirshfeld, Weinhold, and VDD Methods for Charge Analysis. *J. Comput. Chem*. **2004**, *25*, 189-210.

71.     Laidig, K. E.; Bader, R. F., Properties of Atoms in Molecules: Atomic Polarizabilities. *J. Chem. Phys*. **1990**, *93*, 7213-7224.

72.     Wang, J.; Dauter, M.; Alkire, R.; Joachimiak, A.; Dauter, Z., Triclinic Lysozyme at 0.65 Å Resolution. *Acta Crystallogr., Sect. D: Biol. Crystallogr*. **2007**, *63*, 1254-1268.

73.     Schlichting, I.; Berendzen, J.; Chu, K.; Stock, A. M.; Maves, S. A.; Benson, D. E.; Sweet, R. M.; Ringe, D.; Petsko, G. A.; Sligar, S. G., The Catalytic Pathway of Cytochrome P450cam at Atomic Resolution. *Science* **2000**, *287*, 1615-1622.

74.     Rutherford, K.; Le Trong, I.; Stenkamp, R. E.; Parson, W. W., Crystal Structures of Human 108v and 108m Catechol O-Methyltransferase. *J. Mol. Bio*. **2008**, *380*, 120-130.

75.     Labahn, J.; Granzin, J.; Schluckebier, G.; Robinson, D. P.; Jack, W. E.; Schildkraut, I.; Saenger, W., Three-Dimensional Structure of the Adenine-Specific DNA Methyltransferase M.Taq I in Complex with the Cofactor S-Adenosylmethionine. *Proc. Natl. Acad. Sci. U. S. A*. **1994**, *91*, 10957-10961.

76.     Anandakrishnan, R.; Aguilar, B.; Onufriev, A. V., H++ 3.0: Automating pK Prediction and the Preparation of Biomolecular Structures for Atomistic Molecular Modeling and Simulations. *Nucleic Acids Res*. **2012**, *40*, W537-W541.

77.     Gordon, J. C.; Myers, J. B.; Folta, T.; Shoja, V.; Heath, L. S.; Onufriev, A., H++: A Server for Estimating pKas and Adding Missing Hydrogens to Macromolecules. *Nucleic Acids Res*. **2005**, *33*, W368-W371.

78.     Myers, J.; Grothaus, G.; Narayanan, S.; Onufriev, A., A Simple Clustering Algorithm Can Be Accurate Enough for Use in Calculations of pKs in Macromolecules. *Proteins: Struct., Funct., Bioinf*. **2006**, *63*, 928-938.

79.     D.A. Case, J. T. B., R.M. Betz, D.S. Cerutti, T.E. Cheatham, III, T.A. Darden, R.E. Duke, T.J. Giese, H. Gohlke, A.W. Goetz, N. Homeyer, S. Izadi, P. Janowski, J. Kaus, A. Kovalenko, T.S. Lee, S. LeGrand, P. Li, T. Luchko, R. Luo, B. Madej, K.M. Merz, G. Monard, P. Needham, H. Nguyen, H.T. Nguyen, I. Omelyan, A. Onufriev, D.R. Roe, A. Roitberg, R. Salomon-Ferrer, C.L. Simmerling, W. Smith, J. Swails, R.C. Walker, J. Wang, R.M. Wolf, X. Wu, D.M. York and P.A. Kollman Amber 2015, University of California, San Francisco. 2015.

80.     Maier, J. A.; Martinez, C.; Kasavajhala, K.; Wickstrom, L.; Hauser, K. E.; Simmerling, C., Ff14sb: Improving the Accuracy of Protein Side Chain and Backbone Parameters from Ff99sb. *J. Chem. Theory Comput*. **2015**, *11*, 3696-3713.

81.     Hornak, V.; Abel, R.; Okur, A.; Strockbine, B.; Roitberg, A.; Simmerling, C., Comparison of Multiple Amber Force Fields and Development of Improved Protein Backbone Parameters. *Proteins: Struct., Funct., Bioinf*. **2006**, *65*, 712-725.

82.     Wang, J.; Wolf, R. M.; Caldwell, J. W.; Kollman, P. A.; Case, D. A., Development and Testing of a General Amber Force Field. *J. Comput. Chem*. **2004**, *25*, 1157-1174.

83.     Bayly, C. I.; Cieplak, P.; Cornell, W.; Kollman, P. A., A Well-Behaved Electrostatic Potential Based Method Using Charge Restraints for Deriving Atomic Charges: The RESP Model. *J. Phys. Chem.* **1993,** *97*, 10269-10280.

84.     Gordon, M. S.; Schmidt, M. W., Advances in Electronic Structure Theory: GAMESS a Decade Later. *Theory and Applications of Computational Chemistry: the first forty years* **2005**, 1167-1189.

85.     Harihara, P. C.; Pople, J. A., Influence of Polarization Functions on Molecular-Orbital Hydrogenation Energies. *Theor Chim Acta* **1973,** *28*, 213-222.

86.     F. Wang, J.-P. B., P. Cieplak and F.-Y. Dupradeau R.E.D. Python: Object Oriented Programming for Amber Force Fields, Université De Picardie - Jules Verne, Sanford|Burnham Medical Research Institute, Nov. 2013. http://q4md-forcefieldtools.org/REDServer-Development/ (accessed 10/24/16).

87.     Vanquelef, E.; Simon, S.; Marquant, G.; Garcia, E.; Klimerak, G.; Delepine, J. C.; Cieplak, P.; Dupradeau, F.-Y., R.E.D. Server: A Web Service for Deriving RESP and ESP Charges and Building Force Field Libraries for New Molecules and Molecular Fragments. *Nucleic Acids Res.* **2011,** *39*, W511-W517.

88.     Dupradeau, F.-Y.; Pigache, A.; Zaffran, T.; Savineau, C.; Lelong, R.; Grivel, N.; Lelong, D.; Rosanski, W.; Cieplak, P., The R.E.D. Tools: Advances in RESP and ESP Charge Derivation and Force Field Library Building. *Phys. Chem. Chem. Phys.* **2010,** *12*, 7821-7839.

89.     Allnér, O.; Nilsson, L.; Villa, A., Magnesium Ion–Water Coordination and Exchange in Biomolecular Simulations. *J. Chem. Theory Comput.* **2012,** *8*, 1493-1502.

90.     Patra, N.; Ioannidis, E. I.; Kulik, H. J., Computational Investigation of the Interplay of Substrate Positioning and Reactivity in Catechol O-Methyltransferase. *PloS one* **2016,** *11*, e0161868.

91.     Jorgensen, W. L.; Chandrasekhar, J.; Madura, J. D.; Impey, R. W.; Klein, M. L., Comparison of Simple Potential Functions for Simulating Liquid Water. *J. Chem. Phys.* **1983,** *79*, 926-935.

92.     Ryckaert, J.-P.; Ciccotti, G.; Berendsen, H. J. C., Numerical Integration of the Cartesian Equations of Motion of a System with Constraints: Molecular Dynamics of N-Alkanes. *J. Comput. Phys.* **1977,** *23*, 327-341.

93.     Schrodinger, L. L. C., The PyMOL Molecular Graphics System, Version 1.7.4.3. 2010.

94.     Petachem. http://www.petachem.com. (accessed 10/24/16).

95.     Rohrdanz, M. A.; Martins, K. M.; Herbert, J. M., A Long-Range-Corrected Density Functional That Performs Well for Both Ground-State Properties and Time-Dependent Density Functional Theory Excitation Energies, Including Charge-Transfer Excited States. *J. Chem. Phys.* **2009,** *130*, 054112.

96.     Lee, C.; Yang, W.; Parr, R. G., Development of the Colle-Salvetti Correlation-Energy Formula into a Functional of the Electron Density. *Phys. Rev. B* **1988,** *37*, 785-789.

97.     Becke, A. D., Density-Functional Thermochemistry. III. The Role of Exact Exchange. *J. Chem. Phys.* **1993,** *98*, 5648-5652.

98.     Stephens, P. J.; Devlin, F. J.; Chabalowski, C. F.; Frisch, M. J., Ab Initio Calculation of Vibrational Absorption and Circular Dichroism Spectra Using Density Functional Force Fields. *J. Phys. Chem.* **1994,** *98*, 11623-11627.

99.     Ditchfield, R.; Hehre, W. J.; Pople, J. A., Self-Consistent Molecular Orbital Methods. 9. Extended Gaussian-Type Basis for Molecular Orbital Studies of Organic Molecules. *J. Chem. Phys.* **1971,** *54*, 724.

100.    Frisch, M. J.; Pople, J. A.; Binkley, J. S., Self-Consistent Molecular-Orbital Methods .25. Supplementary Functions for Gaussian-Basis Sets. *J. Chem. Phys.* **1984,** *80*, 3265-3269.

101.    Kim, M.-C.; Sim, E.; Burke, K., Understanding and Reducing Errors in Density Functional Calculations. *Phys. Rev. Lett.* **2013,** *111*, 073003.

102.    Rudberg, E., Difficulties in Applying Pure Kohn–Sham Density Functional Theory Electronic Structure Methods to Protein Molecules. *J. Phys. Cond. Matt.* **2012,** *24*, 072202.

103.    Isborn, C. M.; Mar, B. D.; Curchod, B. F. E.; Tavernelli, I.; Martinez, T. J., The Charge Transfer Problem in Density Functional Theory Calculations of Aqueously Solvated Molecules. *J. Phys. Chem. B* **2013,** *117*, 12189-12201.

104.    Cohen, A. J.; Mori-Sánchez, P.; Yang, W., Insights into Current Limitations of Density Functional Theory. *Science* **2008,** *321*, 792-794.

105.    Fleming, A., On a Remarkable Bacteriolytic Element Found in Tissues and Secretions. *Proc. R. Soc. London, Ser. B* **1922,** *93*, 306-317.

106.    Stewart, J. J., Optimization of Parameters for Semiempirical Methods I. Method. *J. Comput. Chem.* **1989,** *10*, 209-220.

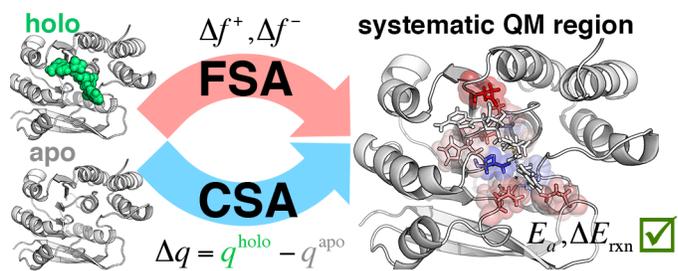